\documentclass[article,aps,nofootinbib]{revtex4}

\usepackage{graphicx}
\usepackage{hyperref}

\def\be{\begin{equation}}
\def\ee{\end{equation}}
\def\ba{\begin{eqnarray}}
\def\ea{\end{eqnarray}}

\newcommand{\bn}{\hat{\bf n}}

\hypersetup{
    colorlinks=true,
    linkcolor=black,
    citecolor=blue,
}

\begin{document}

\title[Searching for Primordial Magnetic Fields with CMB B-modes]{Searching for Primordial Magnetic Fields with CMB B-modes}

\author{Levon Pogosian \& Alex Zucca}
\address{Department of Physics, Simon Fraser University, Burnaby, BC, V5A 1S6, Canada}

\begin{abstract}
Was the primordial universe magnetized? The answer to this question would help explain the origin of micro-Gauss strength magnetic fields observed in galaxies. It is also of fundamental importance in developing a complete theory of the early universe.  While there can be other signatures of cosmological magnetic fields, a signature in the cosmic microwave background (CMB) would prove their primordial origin. The B-mode polarization of CMB is particularly promising in this regard because there are relatively few other sources of B-modes, and because the vortical modes sourced by the primordial magnetic field (PMF) survive diffusion damping up to a small fraction of the Silk length. At present, the Planck temperature and polarization spectra combined with the B-mode spectrum measured by the South Pole Telescope (SPT) constrain the PMF strength to be no more than $\sim 1$ nano-Gauss (nG). Because of the quartic scaling of the CMB anisotropy spectra with the PMF strength, this bound will not change by much even with the significantly better measurements of the B-mode spectrum by the Stage III and Stage IV CMB experiments. On the other hand, being able to tighten the bound well below the $1$ nG threshold is important for ruling out the purely primordial origin (requiring no dynamo action) of galactic fields. Considering Faraday rotation, which converts some of the E-modes into B-modes and scales linearly with the field strength, will help to achieve this goal. As we demonstrate, the upcoming experiments, such as SPT-3G and the Simons Observatory, will be sensitive to fields of $\sim 0.5$ nG strength thanks to the mode-coupling signature induced by Faraday rotation. A future Stage IV ground based experiment or a Space Probe will be capable of probing fields below $0.1$ nG, and would detect a scale-invariant PMF of $0.2$ nG strength without de-lensing or subtracting the galactic rotation measure.
\end{abstract}

\maketitle

\tableofcontents

%
%
%
%

\section{Introduction}

The pre-recombination universe was fully ionized and could sustain a frozen-in magnetic field configuration over many epochs of its evolution \cite{Widrow:2002ud}. There are observational and theoretical arguments favouring the existence of such a primordial magnetic field (PMF)\cite{Grasso:2000wj}. Chief amongst them is the unexplained origin of the galactic magnetic fields, typically of micro-Gauss ($\mu$G) strength and coherent over the extent of the galaxy. Such magnetic fields could be generated through a dynamo mechanism, but it would still require a seed field of a certain minimum strength \cite{Widrow:2011hs}. Moreover, $\mu$G strength fields are observed in high redshift galaxies that are too young to have gone through the number of revolutions necessary for the dynamo to work \cite{Athreya:1998}. Alternatively, the galactic fields could trace their origin to turbulent magnetic field amplification during structure formation at high redshift \cite{Ryu:2008hi,Ryu:2011hu,Schleicher:2012fx,Latif:2012aq,Wagstaff:2013yna}, with the small-scale dynamo converting turbulent energy into magnetic energy. Another possibility is that the supernova explosions in protogalaxies provided the magnetic seed fields that were later amplified by compression, shearing and stochastic motions \cite{Beck:2013gca,Seifried:2013uta}.
A PMF of sufficient strength on appropriate scales could provide the seed, or eliminate the need for dynamical amplification altogether. Since a typical galactic halo is a few kilo-parsec (kpc) in size and forms in a collapse of a few Mega-parsec (Mpc) sized region, the latter possibility requires a PMF of nano-Gauss (nG) present day strength\footnote{It is conventional to use comoving units to describe the strength of cosmological magnetic fields, {\it i.e.} $\mathbf{B}=\mathbf{B}_{\rm today} = a^2\mathbf{B}(a)$, where $a$ is the scale factor normalized to $a=1$ today.} coherent over a comoving Mpc scale in order to compress into a $\mu$G level galactic field.

PMFs are expected to have been produced at some level in the electroweak (EW) and the QCD phase transitions \cite{Vachaspati:1991nm}. Detailed studies of hydromagnetic and magnetohydrodynamic (MHD) turbulence \cite{Brandenburg:2014mwa,Brandenburg:2016odr,Reppin:2017uud,Brandenburg:2017neh} have demonstrated that magnetic power can transfer from smaller to larger scales through both \emph{inverse transfer} and \emph{inverse cascade}, but it is challenging to obtain a coherence scale and the amplitude of the field large enough to explain the galactic fields. For instance, simulations performed in \cite{Brandenburg:2017neh} show that the PMF generated in the EW phase transition can at best reach a strength of $0.3$ nG at a scale of $30$ kpc if the initial field is maximally helical \cite{Brandenburg:2018ptt} and if its initial amplitude is the largest allowed by the big bang nucleosynthesis (BBN) bound \cite{Kernan:1995bz,Grasso:1996kk,Cheng:1996yi,Yamazaki:2012jd}. Another possibility is for the PMF to have been generated during \cite{Turner:1987bw,Ratra:1991bn} and at the end of Inflation \cite{DiazGil:2007dy,DiazGil:2008tf,Adshead:2016iae}, although there it is also challenging to generate a PMF of nG strength. As there are still significant gaps in our understanding of the early universe phase transitions and connecting inflation with particle physics, the possibility of an observationally relevant PMF cannot be ruled out \cite{Durrer:2013pga}. In fact, constraints on the PMF can help us select among different theories of the early universe \cite{Barnaby:2012tk,Long:2013tha}. 

The PMF contributes to CMB anisotropies through metric perturbations and through the Lorentz force felt by baryons in the pre-recombination plasma \cite{Mack:2001gc,Lewis:2004ef,Finelli:2008xh,Paoletti:2008ck,Shaw:2010}. Of particular importance are the vortical perturbations, which lead to B-mode polarization patterns \cite{Subramanian:1997gi, Subramanian:1998fn, Seshadri:2000ky,Subramanian:2002nh,Subramanian:2003sh}. As they contribute to anisotropies on smaller angular scales  ($l \sim 1000$), the magnetic vector modes are not as obscured by the galactic foregrounds as the inflationary B-mode. Another source of magnetic B-modes is Faraday rotation (FR) \cite{Kosowsky:1996yc,Kosowsky:2004zh,Pogosian:2011qv}, which scales linearly with $B$ and, as we will show, offers a very promising probe of the PMF.

Current CMB bounds on the PMF strength are at a few nG level, and have been at a few nG level for many years. The reason for this seemingly slow improvement, despite the significantly better CMB data, is the fact that the CMB anisotropy spectrum induced by the PMF scales as $B^4$. Thus, an order of magnitude stronger constraint on the PMF contribution to the CMB spectra translates into just a factor of $\sim 2$ improvement in the bound on $B$.  Prior to the recent measurements of the B-mode polarization spectrum by the South Pole Telescope (SPT) \cite{Keisler:2015hfa}, the strongest CMB bounds on the PMF strength were derived by the Planck Collaboration \cite{Ade:2015cva} using the 2015 Planck data \cite{Adam:2015rua}. They placed a limit on the PMF strength smoothed over a $1$ Mpc region of $B_{1{\rm Mpc}} < 4.4$ nG at the 95\% confidence level (CL). Comparable bounds were obtained in \cite{Paoletti:2012bb} using the 7-year WMAP data \cite{Larson:2010gs} combined with the high-$l$ temperature anisotropy spectrum from SPT \cite{Keisler:2011aw}.  These constraints assumed a stochastic PMF described by a power-law spectrum with the spectral index treated as a free parameter. The bounds become tighter if one assumes a particular spectral index. For instance, the Planck 2015 bound on the scale-invariant PMF, favoured by the simplest models of inflationary magnetogenesis \cite{Turner:1987bw,Ratra:1991bn}, was $B_{\rm 1Mpc}<2$ nG \cite{Ade:2015cva}.

Recent measurements of the B-mode polarization by POLARBEAR \cite{Ade:2014afa}, BICEP/Keck \cite{Ade:2014xna,Array:2015xqh} and SPT \cite{Keisler:2015hfa} have opened a new opportunity for probing the PMF\cite{Bonvin:2014xia}. The POLARBEAR collaboration placed a bound of $B_{1{\rm Mpc}} < 3.9$ nG  \cite{Ade:2015cao} using only their B-mode spectrum with the cosmological parameters fixed to the best fit LCDM values. Later, a joint fit to the 2015 Planck data and the SPT B-mode spectrum \cite{Keisler:2015hfa}, performed in \cite{Zucca:2016iur}, reduced the Planck-only bound by a factor of two, pushing it to $B_{\rm 1Mpc}<2$ nG at 95\% CL after marginalizing over the spectral index and $B_{\rm 1Mpc}<1.2$ nG for the scale-invariant PMF spectrum. These early results have demonstrated the potential of B-mode measurements for constraining the PMF.  

The bounds mentioned above are based on the analysis of the power spectra of CMB temperature and polarization. Even if the PMF strength $B$ is Gaussian distributed, the PMF stress-energy sourcing the CMB anisotropies is not, as it is quadratic in $B$. Thus, there can be potentially interesting constraints derived from higher order CMB correlations \cite{Brown:2005kr}. Magnetically induced CMB bispectra and trispectra have been studied in \cite{Seshadri:2009sy,Caprini:2009vk,Cai:2010uw,Trivedi:2010gi,Shiraishi:2010yk,Brown:2010jd,Shiraishi:2011dh,Shiraishi:2012rm,Shiraishi:2013wua,Shiraishi:2013vha,Trivedi:2011vt,Trivedi:2013wqa}. It is generally found that they do not improve on the bounds derived from power spectra, with a possible exception of the trispectrum sourced by the {\it passive} modes\footnote{Passive and compensated modes will be discussed in Sec.~\ref{sec:pmf}}. As reported in \cite{Trivedi:2013wqa}, the bound on the trispectrum obtained from the 2013 Planck data \cite{Ade:2013ydc} constrains the amplitude of a scale-invariant PMF to below $0.6$ nG if the energy scale at the time of the PMF generation was $T_B=10^{14}$ GeV. While potentially promising, one should be aware of some factors that may weaken this bound.  The amplitude of the passive mode trispectrum scales as $B^8 (\ln T_B/T_\nu)^4$, where $T_\nu = 1$ MeV is the neutrino decoupling scale, thus the bound on $B$ will weaken if $T_B$, which is unknown, happens to be smaller. Also, most of the contribution comes from the largest angular scales and any increase in the magnetic spectral index will also weaken the bound. Finally, the result in \cite{Trivedi:2013wqa} was based on the variance in the trispectrum calculated by Planck \cite{Ade:2013ydc} without taking into account the PMF contribution, which must be included when constraining the PMF signal.

In addition to sourcing anisotropies, the PMF dumps energy into the plasma, influencing the ionization history of the Universe \cite{Kunze:2014eka, Kunze:2017rsb}. The impact of this effect on the CMB anisotropy spectra offers a potentially more sensitive probe of the PMF than the magnetically induced anisotropies. However, there are remaining uncertainties in modelling the baryon heating due to ambipolar diffusion and decaying magnetic turbulence \cite{Chluba:2015lpa}. The PMF can also induce $\mu$- and $y$-type distortions of the black-body spectrum of the CMB \cite{Jedamzik:1996wp,Jedamzik:1999bm}, which are not well constrained by existing observations but can be promising future probes of the PMF \cite{Kunze:2013uja,Miyamoto:2013oua,Ganc:2014wia,Wagstaff:2015jaa}. Similarly, the PMF affects the thermal Sunyaev-Zeldovich contribution to the CMB through modification of the intergalactic medium evolution \cite{Minoda:2017iob}. It has been suggested that PMF would induce small-scale baryonic density fluctuations at the epoch of cosmic recombination, leading to an inhomogeneous recombination process that alters the heights of the peaks in the CMB spectrum \cite{Jedamzik:2013gua}. This process was recently simulated in \cite{Jedamzik:2018itu} and very strong bounds ($\cal{O}$(0.01-0.05) nG depending on the PMF spectrum) were derived. These results are extremely interesting and deserve follow up investigations.

The search for the PMF is not limited to CMB observations. The magnetic pressure and the diffusion damping of the magnetic fields affect the growth of large scale structure on scales $k > 1$ h/Mpc \cite{Kahniashvili:2010wm,Shaw:2010ea,Kahniashvili:2012dy,Yamazaki:2016sov}. Since non-linearities are important on these scales, accurate modelling of these effects requires numerical simulations of structure formation in the presence of the PMF that do not exist at present. There is also preliminary evidence of magnetic fields coherent over cosmological ($\cal{O}$(Mpc)) distances coming from the non-observation of GeV gamma-ray halos around the TeV blazars \cite{Neronov:1900zz}. Upcoming data, combined with improved modelling of plasma instabilities, should be able to conclusively confirm or deny this possibility \cite{Tashiro:2013ita}. However, even if cosmological scale magnetic fields were discovered, it would not necessarily prove their primordial origin, as they could, in principle, be produced through ejection of magnetic field lines in explosions of highly magnetized stars or via other astrophysical processes \cite{Berezhiani:2003ik,Gnedin:2000ax,Langer:2005gw,Ando:2010ry,Durrive:2015cja}. On the other hand, detecting a magnetic field signature in CMB would provide a conclusive proof of its primordial origin.

The top scientific objective of the next generation CMB polarization experiments is the detection of inflationary gravitational waves \cite{Abazajian:2016yjj}. They will also aim to provide an accurate measurement of the weak lensing contribution to the B-mode spectrum, which happens to be on angular scales where the PMF effects are also important. Their specifications will be well suited for reducing the bound on the PMF strength by an order of magnitude \cite{Yadav:2012uz,De:2013dra,Pogosian:2013dya}. In particular, as we will show, measurements of the FR induced mode-coupling correlations by a Stage IV ground based experiment will be able to probe PMFs of $0.1$ nG strength and finally push the sensitivity well below the critical $1$ nG threshold.

\section{PMF contributions to CMB temperature and polarization anisotropies}
\label{sec:pmf}

Under the ideal magnetohydrodynamic (MHD) approximation, which is known to hold well in the highly conducting primeval plasma of the early universe and on cosmological scales at smaller redshifts, the PMF configuration remains frozen in the plasma. Its strength evolves with time according to $\mathbf{B}(\mathbf{x}, \tau) = \mathbf{B}(\mathbf{x}, \tau_0)/a^2(\tau)$ (see \cite{Subramanian:2015lua} for a review), where $\tau_0$ denotes the present conformal time and $a$ is the scale factor normalized to $a(\tau_0) =1$. Following the convention, we will quote all bounds on the PMF in terms of the ``comoving'' field strength $\mathbf{B}(\mathbf{x}, \tau_0)$. 

We assume that the PMF does not have a homogeneous component, as it would break the isotropy of the Universe and has already been strongly constrained by CMB \cite{Barrow:1997mj,Barrow:1997sy}. Instead, we model the PMF as a statistically isotropic Gaussian distributed random field described by a power spectrum
\begin{equation}
\label{eqn:MagneticPowerSpectrum}
\langle B_i (\mathbf{k}) B_j^*(\mathbf{k}^{\prime}) \rangle = (2 \pi)^3 \delta^{(3)}(\mathbf{k} - \mathbf{k}^{\prime}) P_{ij} P_B(k), 
\end{equation}
where $P_{ij} = \delta_{ij} - \hat{k}_i \hat{k}_j$. In the equation above, we have neglected the parity-odd contribution from the helical component of the PMF. Including the helical component requires introducing additional free parameters and is largely degenerate with the overall normalization, weakening the bounds on the PMF strength by about $25$\% \cite{Ade:2015cva}. Given this, for the sake of simplicity, we omit the helical contribution in the remaining discussion. However, future data can, in principle, measure it through parity-odd CMB polarization spectra \cite{Kahniashvili:2005xe,Kunze:2011bp,Ballardini:2014jta}. 

The magnetic power spectrum is commonly taken to be a power law parametrized as
\ba
P_B(k)  = 
\left\{
    \begin{array}{lr}
      S_0 \, k^{n_B}, &   k<k_D \\
      0, &  k \ge k_D
    \end{array}
\right.
\ea
where $n_B$ is the spectral index and $2\pi/k_D$ is the damping scale below which magnetic fields dissipate due to radiation viscosity \cite{Jedamzik:1996wp, Subramanian:1997gi}.  The spectral index depends on the generation mechanism of the PMF. The originally proposed simple inflationary models of magnetogenesis predicted a scale-invariant spectrum, corresponding to $n_B=-3$, although other values are possible in more complicated models \cite{Bonvin:2013tba}. The PMFs produced in phase transitions generically have $n_B=2$ on large scales probed by CMB \cite{Durrer:2003ja,Jedamzik:2010cy,Kahniashvili:2009qi}. The damping scale depends on the amplitude of the PMF spectrum and the spectral index as \cite{Subramanian:1997gi,Mack:2001gc,Ade:2015cva}
\begin{equation}
{k_D\over \rm{Mpc}^{-1}} = \biggl[ 5.5 \times 10^4 h \biggl( \frac{B_{\lambda}}{nG}\biggr)^{-2} \biggl( \frac{2 \pi}{\lambda / \rm{Mpc}}  \biggr)^{n_B+3} \frac{\Omega_b h^2}{0.022} \biggr]^{\frac{1}{n_B+5}},
\end{equation}
where $\Omega_b$ is the baryon density fraction, $h$ is the reduced Hubble constant, $H_0 = 100 \, h \, \rm{km}/(\rm\rm{s} \,\rm{Mpc})$, and $B_{\lambda}$ is obtained by smoothing the magnetic energy density over a comoving wavelength $\lambda$ using a Gaussian filter:
\ba
B^2_{\lambda} =  \frac{1}{(2 \pi)^3} \int_{0}^{\infty} d^3 k \, e^{-k^2 \lambda^2} P_B(k)
=  \frac{2 S_0}{(2 \pi)^2} \frac{1}{\lambda^{n_B+3}}\Gamma \biggl( \frac{n_B+3}{2}\biggr).
\ea
It is common to take $\lambda = 1 \, \rm{Mpc}$, as it corresponds to the size of a typical region at the time of last scattering that later collapses to form a galactic halo. Another way to quantify the strength of the PMF is in terms of its contribution to the radiation energy density, namely,
\begin{equation}
\mathcal{E}_B = \frac{1}{(2 \pi)^3} \int_0^{k_D} dk \, k^2 P_B(k),
\end{equation}
and defining the effective PMF strength as $B_{\rm{eff}} = \sqrt{8 \pi \mathcal{E}_B}$. Correspondingly, one can define the fractional PMF density as  $\Omega_{B\gamma} = \mathcal{E}_B / \rho_\gamma$ where $\rho_\gamma$ is the total radiation energy density.  The relation between $B_{\rm{eff}}$, $B_{\lambda}$ and $\Omega_{B\gamma}$ is given by \cite{Mack:2001gc,Pogosian:2011qv}
\begin{equation}
B_{\rm{eff}} = \frac{B_{\lambda} (k_D \lambda)^{\frac{n_B+3}{2}}}{\sqrt{\Gamma ((n_B+5)/2)}} = 3.3 \times 10^{3} \sqrt{\Omega_{B\gamma}} \ {\rm nG}.
\label{eq:beff}
\end{equation}
The stress-energy tensor associated with the PMF can be written as \cite{Shaw:2010}
\ba
\label{eqn:PMFEnergyMomentumTensor}
T^{\, \, 0}_{B \,0}(\mathbf{x}, \tau) = - {1 \over 8 \pi a^4} B^2(\mathbf{x}) \equiv -\rho_{\gamma} \Delta_B; \ \ 
T^{\, \,i}_{B \,j}(\mathbf{x}, \tau) = \frac{1}{4 \pi a^4} \biggl( \frac{1}{2} B^2(\mathbf{x}) \delta^i_j - B^{i}(\mathbf{x})B_j(\mathbf{x}) \biggr)
\equiv p_{\gamma}(\Delta_B \delta^i_j + \Pi^{\, \, i}_{B \, j}),
\ea
where $B_i(\mathbf{x}) = B_i (\mathbf{x}, \tau_0)$, $\rho_\gamma$ and $p_\gamma=\rho_\gamma/3$ are the photon density and pressure, $\Delta_B$ is the magnetic contribution to the radiation density contrast and $\Pi^{\, \, i}_{B \, j}$ is the dimensionless anisotropic stress. 

CMB is studied through measurements of four Stokes parameters: the intensity $I$, functions $Q$ and $U$ describing linear polarization, and $V$ which quantifies circular polarization. CMB is not expected to be circularly polarized, and none of the processes discussed in this paper can generate $V$ at an appreciable level, hence we will ignore $V$ from now on. A typical CMB experiment would measure $I$, $Q$ and $U$ at different points on the sky and several frequencies. The primary sources of anisotropies preserve the black body spectrum of the primordial CMB signal, while altering the temperature of the black body along different directions on the sky. For this reason, $I$, $Q$ and $U$ are quantified in terms of fluctuations of the black body temperature in a given line of sight $\bn$. 

Rather than interpreting the data in terms of $Q$ and $U$, which depend on the choice of the coordinate basis, it is beneficial to project them onto a parity-even and a parity-odd basis polarization patterns, referred to as E (gradient) and B (curl) modes \cite{Kamionkowski:1996zd,Seljak:1996gy}. Namely, one can write 
\be
(Q \pm iU)(\bn) = \sum_{lm} (E_{lm}\pm i B_{lm}) _{\pm 2}Y_{lm}(\bn)
\ee
where $_{\pm 2}Y_{lm}(\bn)$ are the spin-$2$ spherical harmonics, and $E_{lm}$ and $B_{lm}$ are the E and B mode multipoles. The E and B mode spectra are obtained by taking ensemble averages of products of the multipole coefficients. For instance, the B-mode spectrum, $C_l^{BB}$, is defined as $\langle B^*_{lm} B_{l'm'} \rangle = \delta_{ll'} \delta_{mm'} C_l^{BB}$. Separating the polarization signal into E and B modes helps to reveal the physical processes responsible for polarizing CMB. Intensity gradients present at last scattering generate parity-even E-mode polarization patterns, while B-modes require sources that have handedness, such as gravitational waves or vorticity in the metric. 

The correspondence between the observed $E$ and $B$ modes and their primordial origin is partially obscured by the galactic foregrounds, primarily the polarized dust and the synchrotron emission, which generate both modes at roughly the same magnitude. They pose a particularly strong challenge for B-modes, since the target primordial signal is extremely faint in comparison. For example, the B-mode spectrum sourced by gravitational waves generated during inflation with tensor-to-scalar ratio $r=0.001$, which is the target of the next generation of CMB experiments, is at least two orders of magnitude weaker than the foreground contribution. Since foregrounds typically depend on frequency, they can be subtracted, in principle, by comparing CMB maps at different frequencies. However, doing it with the above-mentioned accuracy is an extremely challenging ongoing problem. We are specifically interested in B-modes sourced by the PMF.  As we will discuss below, most of the information containing the PMF signatures comes from multipoles in the $500<l<2000$ range, where the EE spectrum peaks. This happens to correspond to the range of scales on which the foregrounds contribute the least. Also, it is sufficient to work with only $10-20$\% of the sky around the galactic poles, where the galactic contamination is smaller.

The PMF stress-energy sources perturbations in the metric, which include vector (vorticity) and tensor (gravitational waves) modes. As we will discuss below, B-modes are primarily sourced by the ``compensated'' vector modes and the ``passive'' tensor modes. In addition, the polarization of CMB is rotated along the line of sight by FR, which is frequency-dependent and introduces a characteristic mode-coupling signature. We briefly review these effects below.

\subsection{Anisotropies from metric fluctuations sourced by a stochastic PMF}

Let us consider a stochastic PMF generated at some early epoch before last scattering. The PMF induced CMB effects are constrained to be small on cosmological scales, hence we can work at linear order in perturbation theory where the PMF is treated as a stiff source with no back-reaction from the metric or matter fluctuations. In Fourier space, the traceless symmetric tensor $\Pi_{B \,j}^{\,i}$ in Eq.~(\ref{eqn:PMFEnergyMomentumTensor}) can be decomposed into scalar (S), vector (V) and tensor (T) components \cite{Kodama:1985bj} as
\begin{equation}
\label{eqn:PiBDecomp}
\Pi_{Bj}^{\,i} (\mathbf{k},\tau) e^{i\, \mathbf{k} \cdot \mathbf{x}}= \Pi_{Bj}^{\, i(S)} + \Pi_{Bj}^{\, i(V)} + \Pi_{Bj}^{\, i(T)},
\end{equation}
with the components given by
\ba
\Pi_{B \,ij}^{(S)} = \Pi_B^{(0)}Q_{ij}^{(0)}; \ \ 
\Pi_{B \,ij}^{(V)} = \Pi_B^{(+1)}Q_{ij}^{(+1)} +  \Pi_B^{(-1)}Q_{ij}^{(-1)} ; \ \
\Pi_{B \,ij}^{(T)} = \Pi_B^{(+2)}Q_{ij}^{(+2)} +  \Pi_B^{(-2)}Q_{ij}^{(-2)}, 
\ea
where $Q_{ij}^{(0)} = -(\hat{k}_i \hat{k}_j - 1/3 \delta_{ij}) \exp{(i\, \mathbf{k} \cdot \mathbf{x})}$, $Q_{ij}^{(\pm 1)} = i \hat{k}_{( \, i} e^{\pm}_{j \, )} \exp{(i \, \mathbf{k} \cdot \mathbf{x})}$ and $Q_{ij}^{(\pm 2)} = e_{ij}^{(\pm 2)} \exp{(i \, \mathbf{k}\cdot\mathbf{x})}$ are, respectively, the scalar, vector and tensor harmonic functions \cite{Kodama:1985bj,Hu:1997hp}. Here, $\mathbf{e}^{(\pm)} = -i/\sqrt{2} (\mathbf{e}^1 \pm i \mathbf{e}^2)$, $\mathbf{e}^{1,2}$ are the unit vectors orthogonal to the wave vector $\mathbf{k}$ and $e_{ij}^{(\pm 2)} = \sqrt{3/2} e_i^{(\pm)} e_j^{(\pm)}$.

At linear order in perturbation theory, the scalar, vector and tensor modes evolve independently. There is a subtlety in setting their initial conditions that stems from the presence of neutrinos. Namely, after neutrinos decouple from photons, which happens at energies below 1 MeV, they free stream and can develop a non-zero anisotropic stress that compensates the stiff anisotropic stress of the PMF. However, prior to their decoupling, neutrinos are bound in a tightly coupled fluid with photons and baryons and are unable to compensate for the magnetic anisotropic stress, which then acts as a source of adiabatic scalar and tensor mode perturbations \cite{Lewis:2004ef}. On scales that remain outside the horizon, the amplitudes of these perturbations grow logarithmically with time until neutrinos decouple and remain constant afterwards. In practice, the initial conditions for perturbations are set on super-horizon scales at some time after the neutrino decoupling. Then, perturbations sourced prior to decoupling set the initial amplitude of the adiabatic mode, which subsequently evolves passively without further influence from the PMF stress-energy. Further, they are assumed to be uncorrelated with the adiabatic fluctuations generated by inflation and are treated as separate \emph{passive} magnetic modes \cite{Lewis:2004ef,Shaw:2010}. After the neutrino decoupling, the PMF stress-energy continues to actively source {\it compensated} perturbations of isocurvature type  \cite{Giovannini:2004aw,Giovannini:2006kc,Finelli:2008xh}, which evolve independently from the adiabatic mode. The adiabatic vector modes decay rapidly when not sourced and can be ignored, thus CMB spectra receive contributions from five magnetic modes in total: a passive and a compensated scalar mode, a passive and a compensated tensor mode, and a compensated vector mode. At linear order, contributions of the different modes to the CMB spectra are uncorrelated and can be computed separately. We will focus only on the two modes that can contribute to the B-mode spectrum at an observationally relevant level: the compensated vector mode and the passive tensor mode\footnote{A detailed discussion of the different modes and their CMB signatures can be found in \cite{Shaw:2010,Ade:2015cva,Zucca:2016iur}}. A publicly available code\footnote{The code can be downloaded at \url{https://alexzucca90.github.io/MagCAMB/}.} for computing PMF sourced CMB spectra was developed in \cite{Zucca:2016iur}  based on CAMB \cite{Lewis:1999bs} and an earlier code written by R.~Shaw \cite{Shaw:2010}.

The PMF acts as an active source of vector perturbations with the vector component of the magnetic anisotropic stress, $\Pi_B^{(\pm 1)}$, being compensated by the anisotropic stress of neutrinos, $\Pi_{\nu}^{(\pm 1)}$. The compensated tensor mode has a negligibly small amplitude and can be ignored. However, by the time of neutrino decoupling, the PMF sources an adiabatic tensor mode of amplitude \cite{Shaw:2010}
\begin{equation}
H^{\pm 2} \approx R_{\gamma} \Pi_B^{(\pm 2)} \biggl[ \ln(\tau_{\nu} / \tau_B) + \biggl(\frac{5}{8 R_{\nu}} -1 \biggr) \biggr],
\end{equation}
where $\tau_{\nu}$ and $\tau_B$ are the conformal times at the epochs of neutrino decoupling and the PMF generation, and $R_{\gamma}$ and $R_{\nu}$ are the photon and neutrino fractions of the total radiation density. 

\begin{figure}[tbp]
\centering
\includegraphics[width=0.32\textwidth]{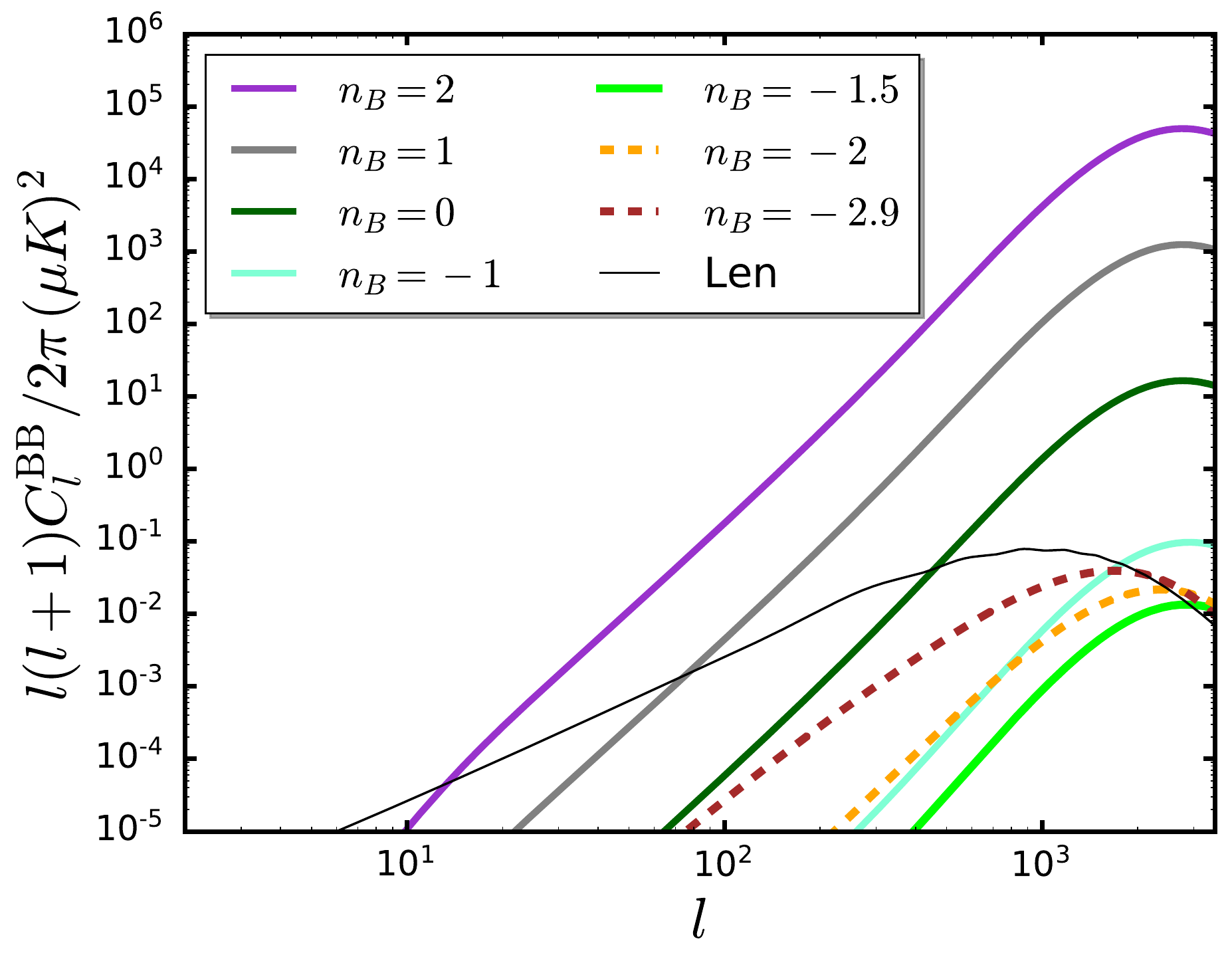}
\includegraphics[width=0.32\textwidth]{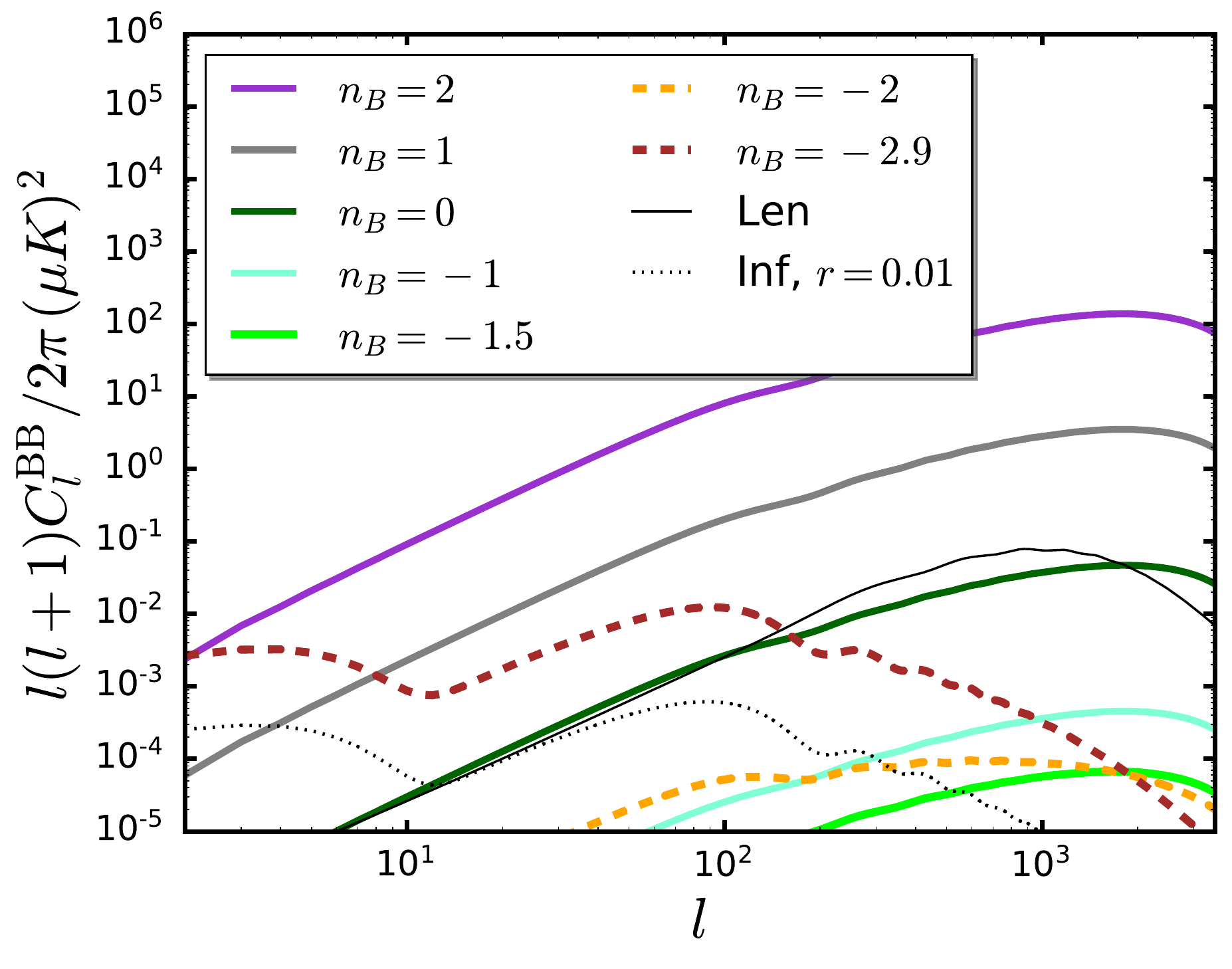}
\includegraphics[width=0.32\textwidth]{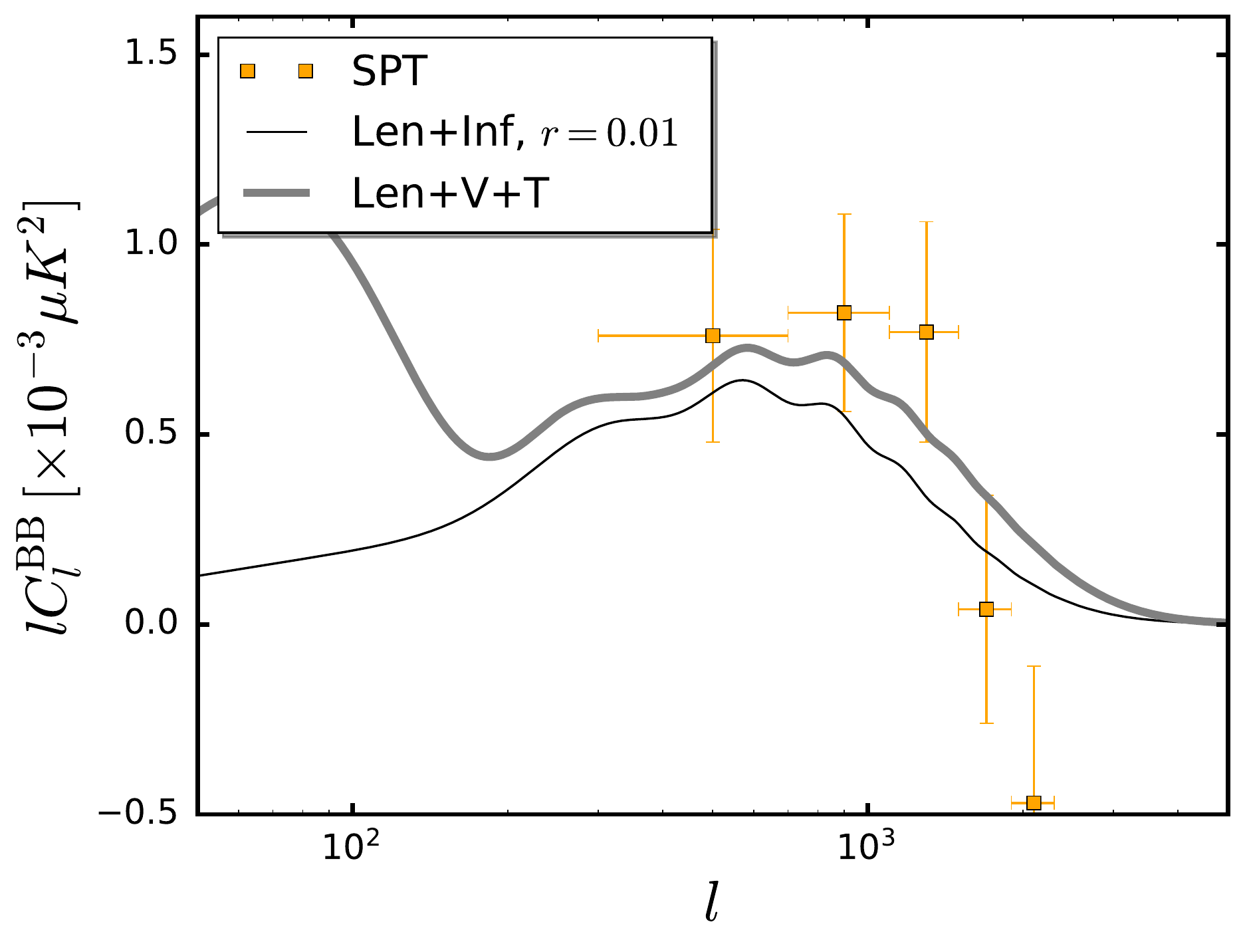}
\caption{\label{fig:VecBmode} The B-mode spectrum from the PMF vector compensated mode (left panel) and tensor passive mode (center panel) for $B_{1 \, \mathrm{Mpc}} = 2 \, \mathrm{nG}$ and different values of the spectral index $n_B$. The epoch of generation of the PMF is $\beta =\log_{10} (\tau_{\nu} / \tau_{B}) = 17$. The thin black solid line shows the lensing contribution and the thin dotted line shows the inflationary tensor mode with $r=0.01$. Right panel: the B-mode spectrum from the PMF vector and tensor modes added to the lensing contribution (solid black line) for $B_{1 \, \rm{Mpc}}=2$ nG, $n_B=-2.9$ and  $\beta =\log_{10} (\tau_{\nu} / \tau_{B}) = 17$, compared to the lensing BB combined with the inflationary contribution with $r=0.01$. The SPT B-mode combined measurements are shown with oranges squares.}
\end{figure}

In Fig.~\ref{fig:VecBmode} (left and centre panels), we show the compensated vector and the passive tensor  modes contributions to the B-mode spectrum (BB) for $B_{1 \, \rm{Mpc}} = 2$ nG, $\beta=\log_{10} (\tau_{\nu}/\tau_B) = 17$ and different values of the spectral index. Note the change in the dependence on $n_B$ that occurs at $n_B = -1.5$. In the $-3<n_B<-1.5$ range, the shapes of the CMB spectra are determined by a Fourier spectrum that scales as $k^{2 n_B + 6}$ \cite{Shaw:2010}. Thus, in that regime, an increase in $n_B$ causes a shift of power from lower to higher $l$. This reduces the CMB anisotropy power on scales within the observational window. In contrast, for $n_B \ge -1.5$, the CMB spectra are independent on $n_B$ because the Fourier spectrum is cutoff dominated. In that case, the amplitude of the spectrum increases with $n_B$, thus larger values of $n_B$ lead to more CMB power for the same PMF strength on $1$ Mpc scale. 

The vector mode contribution to BB, being independent of $\tau_B$, offers a less model-dependent constraint on the PMF strength. Furthermore, it contributes to scales on which foregrounds are less of a problem and where the lensing contribution to the B-mode spectrum has already been measured. In the future, when measurements at lower $l$ become available, simultaneous constraints on the tensor and vector modes will allow to constrain $\tau_B$, which is of fundamental interest. 

The contribution of the magnetic vector mode to the B-mode polarization power spectrum can be well constrained by the current and future CMB experiments capable of detecting the B-modes from weak lensing. It is clear from the right panel of Fig.~\ref{fig:VecBmode} that the high-$l$ measurements of the B-mode polarization performed by SPT can place competitive bounds on the amplitude of the PMF. As shown in \cite{Zucca:2016iur}, they significantly reduce the upper bound obtained from Planck.

\subsection{Mode-coupling correlations induced by Faraday rotation}
\label{fr_basics}

The polarization vector of CMB photons passing through magnetized plasma is rotated by an angle \cite{Kosowsky:1996yc,Harari:1996ac}
\begin{equation}
\alpha(\hat{\bf n}) = \frac{3c^2 \nu^{-2}}{{16 \pi^2 e}}
\int \dot{\tau} \ {\bf B} \cdot d{\bf l} 
=c^2 \nu^{-2} \ {\rm RM}(\hat{\bf n}),
\label{alpha-FR}
\end{equation}
where $\dot{\tau}$ is the differential optical depth, 
$\nu$ is the frequency at which CMB is observed, ${\bf B}$ is the comoving magnetic field, and $d{\bf l}$ is the comoving length element along the photon trajectory. The rotation measure (RM) is a commonly used frequency independent characteristic of the strength of FR. Eq.~(\ref{alpha-FR}) implies that a significant RM can be produced by a small PMF over a large distance, which is the case at recombination, or by a larger magnetic field over a smaller path, which is the case inside our galaxy. The latter acts as noise in the estimates of the RM due to a PMF and was studied in detail in \cite{De:2013dra}. The main contribution of the PMF to FR happened just after the polarization was generated at the time of last scattering, when the mean ionized fraction was still high. Subsequently, additional FR was produced in ionized regions along the line of sight that contain magnetic fields, such as clusters of galaxies and our own galaxy. We will not discuss FR from clusters, as they are expected to induce rotations of CMB polarization on relatively small scales. On the other hand, the Milky Way contribution to the RM near galactic poles can look very similar to that of a scale-invariant PMF \cite{De:2013dra}.

When the polarization is rotated, the two Stokes parameters transform as
\begin{equation}
Q(\nu) + iU(\nu) =(Q^{(0)} + iU^{(0)}) \exp(2i\alpha(\nu)) \ ,
\label{qu-rotation}
\end{equation}
where $Q^{(0)}$ and $U^{(0)}$ are the Stokes parameters at last scattering. Since the FR angle falls off as $1/\nu^2$, $Q^{(0)}$ and $U^{(0)}$ can also be taken to be the Stokes parameters measured at a very high frequency. In theory, it is possible to extract the FR angle by taking maps of $Q$ and $U$ at different frequencies and using Eq.~(\ref{qu-rotation}) 
to solve for the rotation in each pixel. Each additional frequency channel provides a separate measurement of $\alpha(\nu)$ thus reducing the error bar on the measurement of ${\rm RM}(\bn)$. Such a direct measurement of FR would be interesting to attempt, but could be challenging given that $Q$ and $U$ maps require significant processing to subtract the foreground contributions.

Another way to extract the rotation angle is based on quadratic estimators utilizing mode-coupling correlations between E and B modes induced by FR \cite{Kamionkowski:2008fp,Yadav:2009eb,Gluscevic:2009mm,Gluscevic:2012me}. It is similar to the procedure used to isolate the weak lensing contribution to CMB anisotropies \cite{Hu:2001kj}. Unlike a direct extraction of FR from Eq.~(\ref{qu-rotation}), the quadratic estimator method does not utilize the frequency dependence. It is statistical in nature and formally involves summing over all pixels of $Q$ and $U$ in order to reconstruct $\alpha$ in a given direction on the sky. 

In the small rotation angle approximation, which is well-justified since the PMF is constrained to be very small, the relation between the spherical expansion coefficients of the E, B and $\alpha$ fields can be written as
\be
B_{lm}=2\sum_{LM}\sum_{l' m'}\alpha_{LM} E_{l' m'} 
\xi_{lml'm'}^{LM}H_{ll'}^L \ ,
\label{eq:blm}
\ee
where $\xi_{lml'm_2}^{LM}$ and $H_{ll'}^L$ are related to Wigner $3$-$j$ symbols via \cite{Gluscevic:2012me}
\ba
\nonumber
\xi_{lml'm'}^{LM} &\equiv& (-1)^m \sqrt{ (2l+1)(2L+1)(2l'+1) \over 4\pi}
\left(
\begin{array}{ccc}
l  & L  & l'  \\
-m  & M  & m'    
\end{array}
\right) \\
H_{ll'}^L &\equiv& 
\left(
\begin{array}{ccc}
l  & L  & l'  \\
2  & 0  & -2    
\end{array}
\right) \ ,
\ea
with the summation restricted to even $L+l'+l$. Eq.~(\ref{eq:blm}) assumes ``instant last scattering'', namely, that all of the E-mode polarization was generated instantly at some early time, and subsequently rotated. In reality, during last scattering, the polarization is rotated while it is being produced. This approximation has been investigated in great detail in \cite{Pogosian:2011qv}, where the exact radiation transfer equations in the presence of magnetic fields were solved. It was shown that corrections introduced by taking into account the width of the last scattering are not negligible, but not significant enough to warrant complicating the discussion of the observability of the FR effects.

Eq.~(\ref{eq:blm}) implies correlations between multipoles of E and B modes. Since the primordial T and E are correlated, FR also correlates T and B. The quadratic estimator method assumes the primordial $E$ mode to be a statistically isotropic Gaussian random field characterized by a known power spectrum, so that $\langle E_{l m}^* E_{l'm'} \rangle =  \delta_{ll'} \delta_{mm'} C_l^{EE}$. Then, given a CMB polarization map, one can calculate a quantity
\be
[{\hat \alpha}_{BE,LM}]_{ll'} = {2\pi \sum_{mm'} B_{lm}E^*_{l'm'}\xi_{lml'm'}^{LM} \over (2l+1)(2l'+1)C_l^{EE}H_{ll'}^L}
\label{alphallpr}
\ee
which is an unbiased estimator of $\alpha_{LM}$ \cite{Kamionkowski:2008fp} from a given pair of multipoles $ll'$. 
The optimal estimate of $\alpha_{LM}$ is obtained by summing over all available $ll'$ pairs weighted appropriately to minimize the variance. An analogous estimator can be constructed from products of T and B. The variance is further reduced when combinations of E, B and T from different frequency channels are combined. In such a case, one works with the estimator of RM defined, in the case of EB, as \cite{Pogosian:2013dya} 
\be
[{\hat r}_{B^iE^j,LM}]_{ll'} = c^{-2} \nu^{2}_i [{\hat \alpha}_{B^iE^j,LM}]_{ll'} \ ,
\ee
where $i$ and $j$ label different frequency channels. 

The variance in ${\hat r}_{LM}$, for a statistically isotropic RM, is defined as
$\langle {\hat r}^*_{LM} {\hat r}_{L'M'} \rangle =\delta_{LL'} \delta_{MM'} [C_L^{\rm RM}+\sigma^2_{{\rm RM},L}]$,
where $C_L^{\rm RM}$ is the RM power spectrum that receives contributions from the PMF and the galaxy, while $\sigma^2_{{\rm RM},L}$ is the combined variance of individual estimators $[{\hat r}_{B^iE^j,LM}]_{ll'}$. Using notation similar to that of \cite{Gluscevic:2009mm}, the variance can be written as
\be
\sigma^{-2}_{{\rm RM},L} = \sum_{l, l' \ge l} G_{l l'} 
\sum_{A,A'} [({\cal C}^{l l'})^{-1}]_{AA'} \ 
Z_{l l'}^A Z_{l l'}^{A'} 
\ ,
\label{eq:ebnoise}
\ee
where the sum is restricted to even $l+l'+L$, $G_{l l'} \equiv (2l+1)(2l'+1) (H^L_{l l'})^2 / \pi$, $A$ and $A'$ label all relevant quadratic combinations of E, B and T, 
\ba
Z_{l l'}^{X^iB^j} = c^2\nu_j^{-2} W^{ij}_{l l'} C^{XE}_l  , \\
Z_{l l'}^{B^iX^j} = c^2\nu_i^{-2} W^{ij}_{l l'} C^{EX}_{l'} ,
\ea
with $X$ denoting either $T$ or $E$, and $W^{ij}_{l l'} \equiv \exp[-(l^2+l'^{2}) \theta^2_{ij}/16\ln 2]$ accounts for the finite width of the beam. We take $\theta_{ij} = \max[{\theta^i_{\rm fwhm},\theta^j_{\rm fwhm}}]$, where $\theta^i_{\rm fwhm}$ is the full-width-at-half maximum (FWHM) of the Gaussian beam of the $i$-th channel. The covariance matrix elements, $[{\cal C}^{l l'}]_{AA'}$, are
\be
[{\cal C}^{l l'}]_{X^iB^j,Y^kB^n}= {\tilde C}^{X^iY^k}_l {\tilde C}^{B^jB^n}_{l'} ; \ 
[{\cal C}^{l l'}]_{B^iX^j,B^kY^n}= {\tilde C}^{B^iB^k}_l {\tilde C}^{X^jY^n}_{l'}
\ee
with $X$ and $Y$ standing for either E or T, and
\be
{\tilde C}^{X^iY^j}_l \equiv C^{XY, {\rm prim}}_l+f_{\rm L}C^{XY, {\rm WL}}_l+ \delta_{X^iY^j} \sigma^2_{P,i} \ ,
\label{clvariance}
\ee
is the measured spectrum, that includes the primordial contribution $C^{XY, {\rm prim}}_l$, the WL contribution $C^{XY, {\rm WL}}_l$, and the detector noise $\sigma^2_{P,i}$ which is taken to be uncorrelated between different maps. The de-lensing fraction $f_{\rm L}$ is introduced to account for the partial subtraction of the WL contribution. According to \cite{Seljak:2003pn}, the quadratic estimator method of \cite{Hu:2001kj} can reduce the WL contribution to ${\tilde C}_l^{BB}$ by a factor of $7$ (implying $f_{\rm L} =0.14$), with iterative methods promising a further reduction \cite{Seljak:2003pn}.

The signal-to-noise ratio of the detection of the primordial RM spectrum $C_L^{\rm RM,PMF}$ is given by
\be
\left( S \over N \right)^2 = \sum_{L=1}^{L_{max}} {(f_{\rm sky}/2) (2L+1) [C_L^{\rm RM,PMF}]^2 \over [C_L^{\rm RM,PMF} + f_{\rm G}C_L^{\rm RM,G}+\sigma^2_{{\rm RM},L}]^2} \ ,
\label{eq:ebsnrP}
\ee
where $f_{\rm G}$ is the fraction of the galactic RM spectrum that contributes to the total measured signal. Estimates of $C_L^{\rm RM,G}$ for different sky cuts were obtained in \cite{De:2013dra} based on the galactic RM map of Oppermann et al \cite{Oppermann:2011td}. The RM map \cite{Oppermann:2011td} was produced using a compilation of FR of extragalactic radio sources. They found that the RM is the strongest along the galactic plane, falling off as one moves towards the poles and becoming latitude independent at latitudes higher than $80^\circ$. Near the galactic poles, the RM is $\sim 20$ rad/m$^2$ and has a nearly scale-invariant anisotropy spectrum for $L>200$. It was found that, because of the strong RM away from the poles, it is optimal to use $f_{\rm sky} \sim 0.2$ for constraining the PMF.

The quadratic estimator works for the monopole ($L=0$) and the dipole ($L=1$) of the FR field which, in principle, should not be ignored. However, the FR angle is not expected to have a monopole, since it would imply a non-zero magnetic charge enclosed by the CMB surface, while the dipole, corresponding to a uniform magnetic field, is constrained to be negligible for the PMF and is relatively unimportant for the galactic RM.

The variance of the rotation measure estimator $\sigma_{{\rm RM},L}$ is the smallest for large scale rotations and increases steadily with increasing $L$. This makes physical sense -- a rotation that is uniform over a large part of the sky will couple many $ll'$ pairs in exactly the same way, boosting the signal-to-noise of its detection. Thus, while one wants to measure E and B modes at $l \sim 1000$, one is actually probing rotations at $L < 100$ \cite{Pogosian:2013dya}. For this reason, the mode-coupling estimator is generally more sensitive to PMFs that have a large fraction of their power on large scales. 

For a scale-invariant PMF, the dependence of the signal-to-noise on the survey parameters can be approximated by \cite{Pogosian:2013dya} 
\be
{S \over N} \approx \left( B_{\lambda} \over 0.1 \ {\rm nG} \right) \left( f_{\rm sky} \over 0.2 \right)^{1/2} \left(28' - \theta_{\rm fwhm} \over 10' \right) \left(95 {\rm GHz} \over \nu \right)^2 \left(1 \mu \text{K-arcmin} \over \sigma_{P} \right) \ ,
\label{eq:sn-fr}
\ee
which works reasonably well for experiments capable of detecting sub-nG level fields. One can deduce the expected bound on the PMF strength from (\ref{eq:sn-fr}) by setting the $S/N$ according to the desired significance level, {\it e.g.} $S/N=1$ for the 1$\sigma$ bound,  or $S/N=2$ for 2$\sigma$, {\it etc}, and solving for $B_{\lambda}$. The reader should be cautioned that this expression does not account for the weak lensing contribution and the galactic RM and hence should be taken only as a very approximate ``in principle'' limit achievable with a given experiment. In the forecasts presented in the next Section, all bounds are calculated numerically using the exact expression (\ref{eq:ebsnrP}).

\begin{figure}[tbp]
\centering
\includegraphics[width=0.5\textwidth]{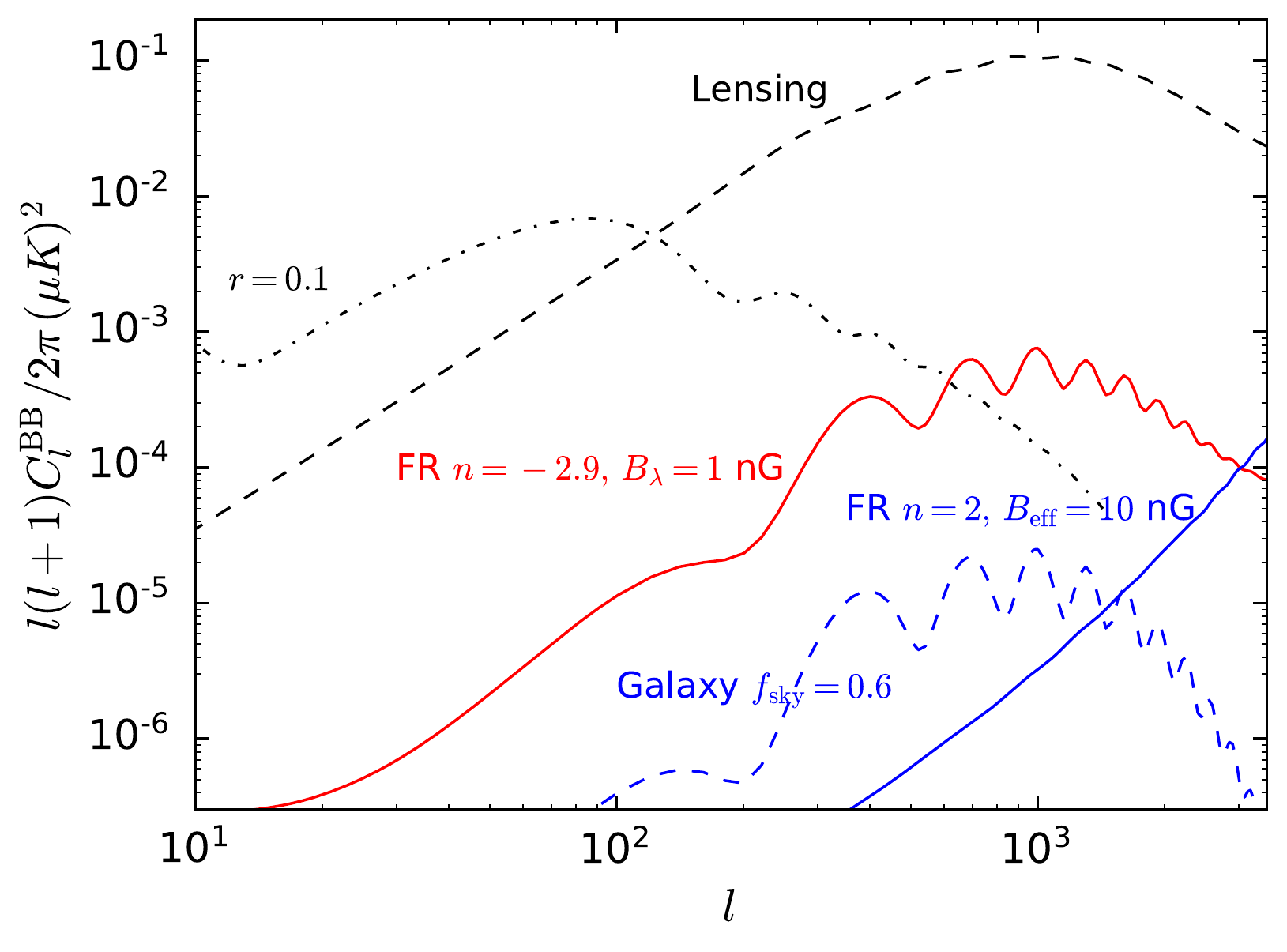}
\caption{\label{fig:bb_fr} The CMB B-mode spectrum from Faraday rotation (FR) sourced at $70$~GHz by a nearly scale-invariant PMF with $n_B=-2.9$ and $B_\lambda = 1$ nG (red solid line), a PMF with $n_B=2$ and $B_{\rm eff}=10$ nG (blue solid line), and the FR due to the galactic magnetic field with Planck's sky mask ($f_{\rm sky}=0.6$). Note that the amplitude of the FR BB scales with the frequency as $\nu^{-4}$. Shown for comparison are contributions from inflationary gravitational waves with $r=0.1$ (black dash-dot line) and from weak gravitational lensing of CMB by the large scale structure (black long-dash).}
\end{figure}

In addition to the mode-coupling correlations of EB and TB type, FR contributes to the B-mode polarization spectrum, $C_l^{BB}$ \cite{Kosowsky:2004zh,Kahniashvili:2008hx}. A detailed discussion of the FR induced B-mode spectrum can be found in \cite{Pogosian:2011qv}. As shown in \cite{De:2013dra}, for scale-invariant PMF, the FR signatures are always less visible in the B-mode spectrum compared to the mode-coupling quadratic estimator. This is due to two primary factors: a) the mode-coupling estimator is linear in the small rotation angle $\alpha$, while BB scales as $\alpha^2$, and b) in the mode-coupling estimator (\ref{alphallpr}), many $ll'$ multipole pairs of E and B fields probe the same $L$ mode of the rotation field. The mode-coupling estimator is primarily sensitive to large scale rotations and, for that reason, it is more sensitive to near-scale-invariant PMFs and less so to causally generated PMF, which have blue spectra with $n_B=2$. In the latter case, most of the FR rotation happens on very small scales and the S/N of detecting FR in BB is actually higher than that in the quadratic estimator \cite{Yadav:2012uz}. 

Fig.~\ref{fig:bb_fr} shows the FR induced BB spectra at $70$ GHz for the scale-invariant field of $1$ nG strength and a causal field with $B_{\rm eff}=10$ nG. Also shown is the BB spectrum from the galactic FR obtained in \cite{De:2013dra} using the rotation measure map of \cite{Oppermann:2011td} after applying the Planck galactic cut. As one can see, the FR sourced BB is always well below the weak lensing signal. Note that the amplitude of the FR sourced B-mode spectrum scales as $\nu^{-4}$, and the relatively foreground-free CMB frequencies are in the 90-150 GHz range. Thus, overall, the BB sourced by primordial or the galactic FR is not expected to be significant and we will not discuss it further.

\section{Current and future B-mode bounds on primordial magnetic fields}

CMB polarization was first detected by DASI in 2002 \cite{Kovac:2002fg}, followed shortly by WMAP \cite{Kogut:2003et,Page:2006hz}. In the subsequent decade, the E-mode polarization measurements steadily improved, culminating in the high-resolution EE spectra obtained by Planck \cite{Adam:2015rua}. Until recently, only upper bounds on the B-mode spectrum were available. However, thanks to the pioneering work by the POLARBEAR \cite{Ade:2014afa}, BICEP/Keck \cite{Ade:2014xna,Array:2015xqh} and SPT \cite{Keisler:2015hfa} collaborations, cosmology has entered the era of precision B-mode science. The ongoing and future experiments will improve the accuracy and resolution of B-mode measurements by orders of magnitude. In what follows, we will review the current PMF bounds derived from B-modes and present forecasts for the upcoming and future CMB experiments. We will consider both ways in which constraints on the PMF can be derived from CMB polarization, namely, the B-mode spectrum and the mode-coupling correlations induced by FR.

\subsection{Current bounds from the B-mode spectrum and Faraday rotation}

\begin{figure}[tbp]
\centering
\includegraphics[width=0.32\textwidth]{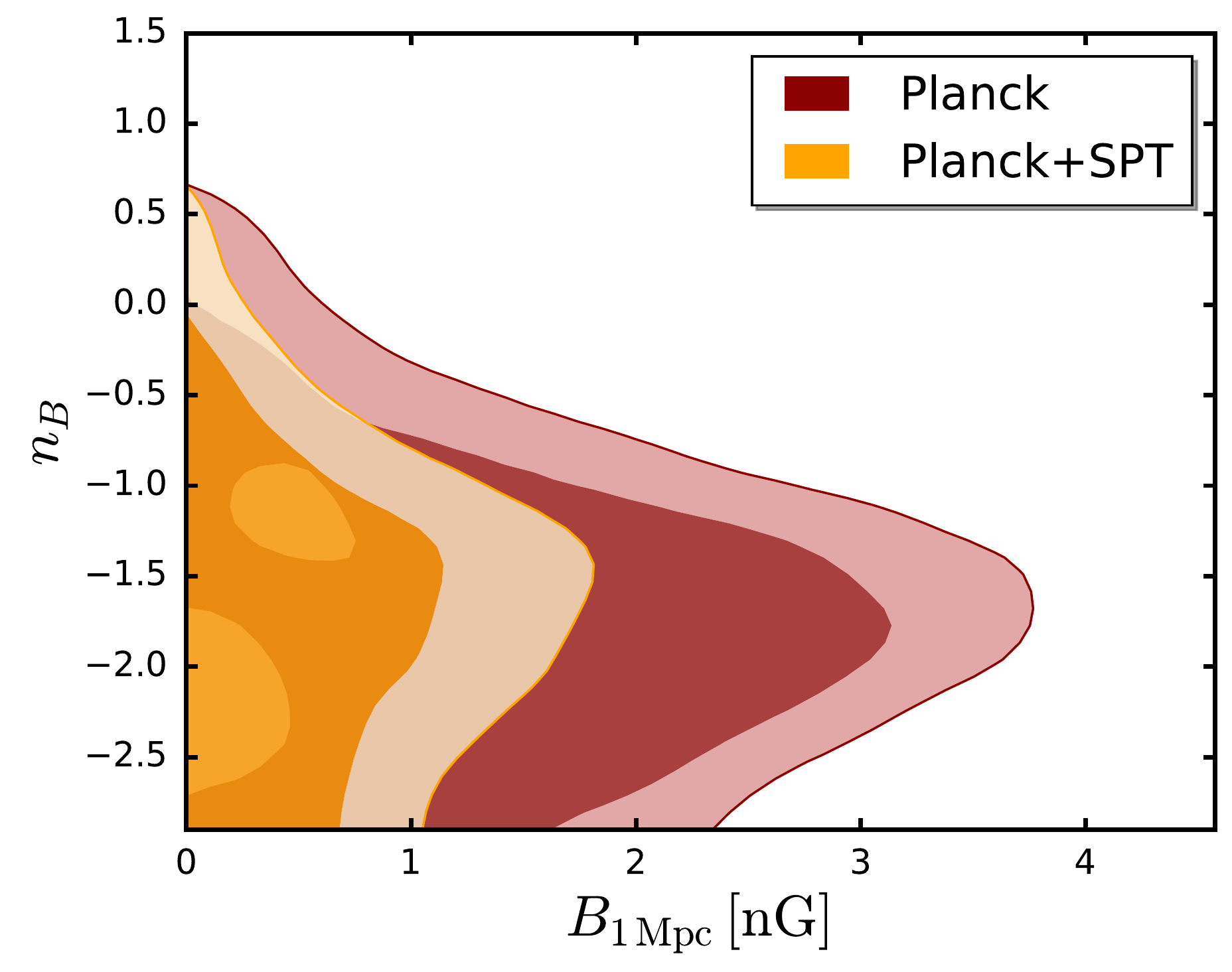}
\includegraphics[width=0.32\textwidth]{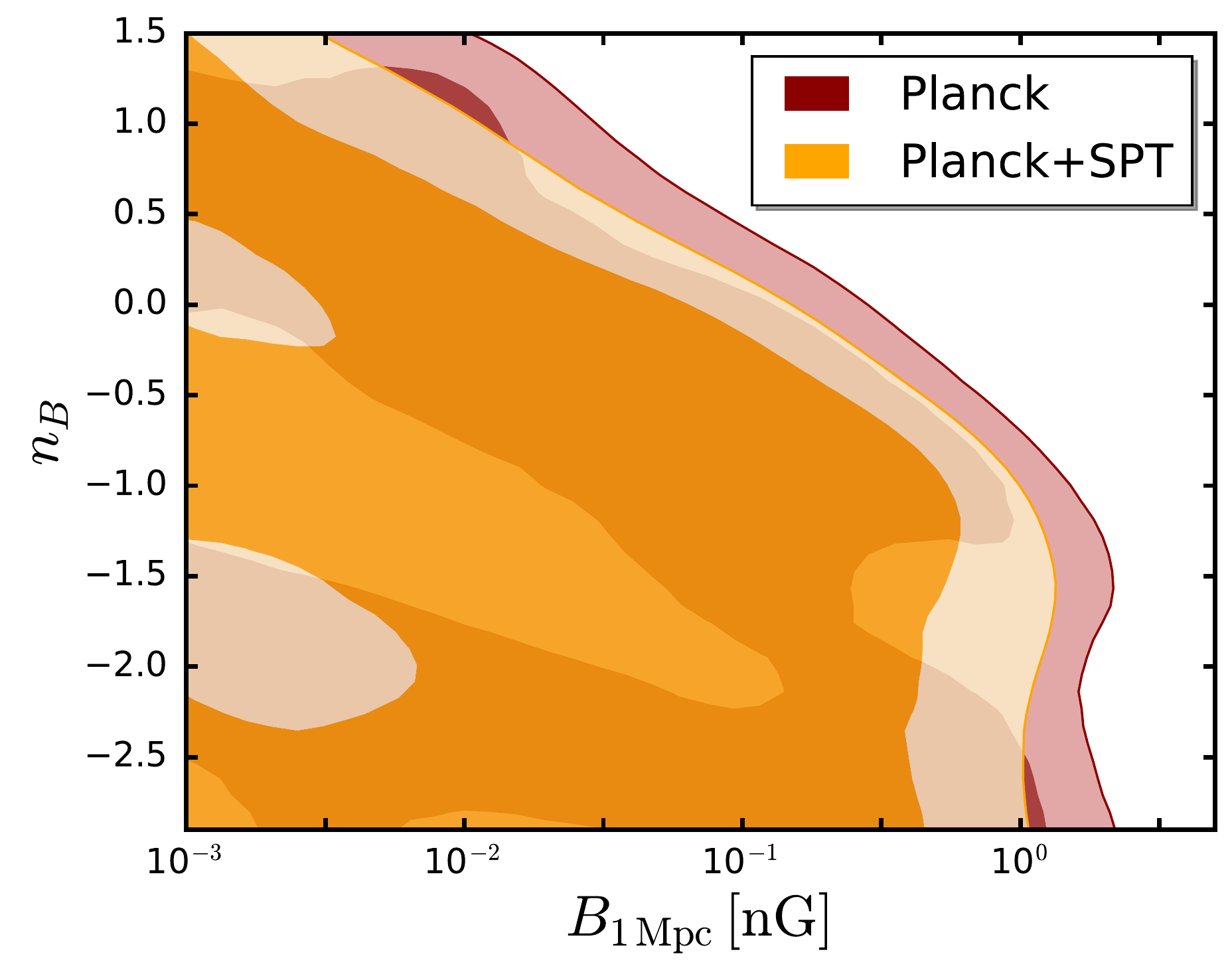}
\includegraphics[width=0.32\textwidth]{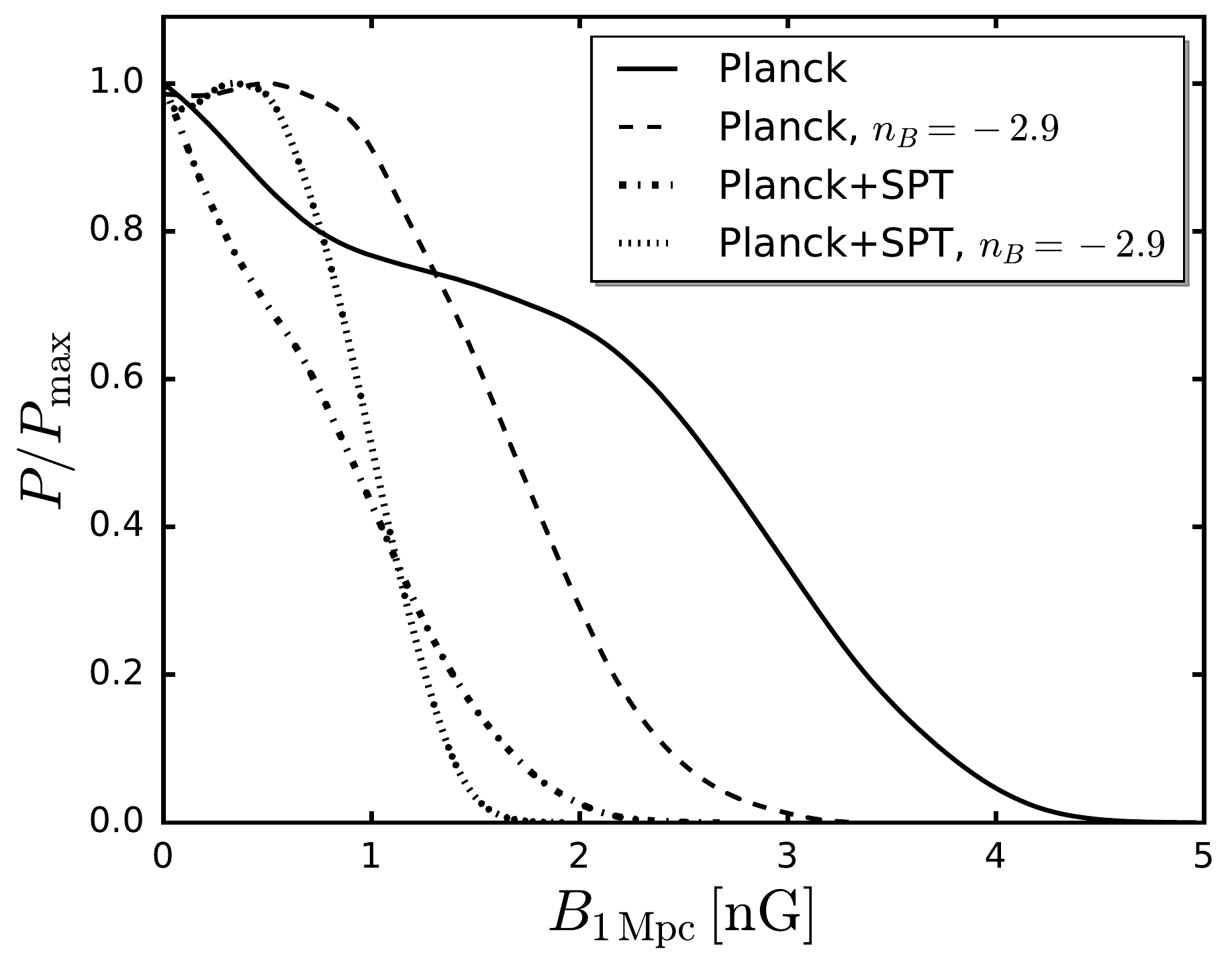}
\caption{\label{fig:PlanckContours} Left panel: the joint probability for the magnetic amplitude $B_{1 \, \rm{Mpc}}$ and the magnetic spectral index $n_B$ using a uniform prior on $B_{1 \, \rm{Mpc}}$.  Center panel: the joint probability for  $B_{1 \, \rm{Mpc}}$ and  $n_B$ using a uniform prior on $\log_{10}(B_{1 \, \rm{Mpc}}/\mathrm{nG})$.  The two shaded regions represent the 68\% CL and 95\% CL, respectively.  The apparent bound on $n_B$ in the left panel disappears, as expected, when the logarithmic prior is used. Right panel: the marginalized PDFs for the magnetic amplitude $B_{1 \, \rm{Mpc}}$ from  Planck and the combination of Planck and SPT. Also shown are PDFs in the nearly scale-invariant case, $n_B=-2.9$.}
\end{figure}

Prior to the B-mode data becoming available, CMB bounds on the PMF were derived from the available measurements of TT, TE and EE \cite{Paoletti:2012bb,Ade:2015cva}. These spectra are dominated by the primary CMB anisotropies, with the PMF contribution being relatively small and partially degenerate with changes in cosmological parameters. The availability of the B-mode spectra significantly improved the bounds. This was first demonstrated by the POLARBEAR collaboration, who placed a bound of $B_{1{\rm Mpc}} < 3.9$ nG \cite{Ade:2015cao}, comparable to the one from Planck, using only their B-mode spectrum. One should note that Planck data was used indirectly to determine the cosmological parameters assumed in \cite{Ade:2015cao}, however the PMF contributions to TT, TE and EE were not used in deriving the constraint. 

A comprehensive analysis was performed by Zucca et al \cite{Zucca:2016iur}, who considered the PMF contributions to all CMB spectra and obtained constraints based on the Planck TT, TE and EE \cite{Aghanim:2015xee} in combination with the B-mode spectrum measured by SPT \cite{Keisler:2015hfa}. Fig.~\ref{fig:PlanckContours}, based on the analysis in \cite{Zucca:2016iur}, allows one to compare the PMF bounds obtained from the Planck data alone, and after adding the BB measurements. The PMF bounds obtained by the Planck collaboration in \cite{Ade:2015cva} are broadly consistent with \cite{Zucca:2016iur}, with minor differences that have been addressed in \cite{Zucca:2016iur}.

To obtain the distributions in Fig.~\ref{fig:PlanckContours}, in addition to the standard cosmological parameters, three PMF parameters were varied: the amplitude $B_{1 \rm{Mpc}}$, the spectral index $n_B$, and the PMF generation time parameter $\beta = \log_{10}(\tau_{\nu} / \tau_B)$. The left and the centre panels of Fig.~\ref{fig:PlanckContours} show the joint probability obtained in \cite{Zucca:2016iur} for the magnetic amplitude $B_{1 \, \rm{Mpc}}$ and the magnetic index $n_B$ from Planck alone and after combining Planck with SPT. One can see that the bound on $B_{1 \, \rm{Mpc}}$ becomes weaker with increasing $n_B$ in the $-3<n_B<-1.5$ range. However, the bound becomes stronger when $n_B>-1.5$. This is due the qualitative change in the dependence of the CMB spectra on $n_B$ that occurs at $n_B=-1.5$.  Namely, as discussed in \cite{Ade:2015cva, Zucca:2016iur} and can be seen from Fig.~\ref{fig:VecBmode}, for $-3<n_B<-1.5$, an increase in $n_B$ results in a shift of power from lower to higher $l$, reducing the signal inside the observational window, and thus allowing for larger PMF strengths. In contrast, when $n_B>-1.5$, the CMB spectra are cutoff dominated and no longer change in shape, with larger $n_B$ giving larger amplitudes for the same PMF strength, leading to tighter constraints on $B_{1 \, \rm{Mpc}}$.

Since there was no detection of the PMF amplitude, the magnetic spectral index $n_B$ and the PMF generation epoch parameter $\beta$ are unconstrained. To see this, one must choose appropriate priors on $B_{1 \, \rm{Mpc}}$. Fig.~\ref{fig:PlanckContours} separately shows the cases with a uniform (left panel) and the logarithmic (center panel) prior on $B_{1 \, \rm{Mpc}}$. One can see an apparent upper bound on $n_B$ in the case of the uniform prior, which is clearly unphysical and is due to the fact that extremely small values of $B_{1 \, \rm{Mpc}}$ have effectively zero prior probability. The upper bound on $n_B$ disappears if one uses a logarithmic prior on $B_{1 \, \rm{Mpc}}$ which treats values of different orders of magnitude with equal probability. Incidentally, the choice of the prior does not affect the upper bound on $B_{1 \, \rm{Mpc}}$.

There is a potential degeneracy between the tensor-to-scalar ratio $r$ and the time of the generation of the PMF parameter $\beta$, since, for nearly scale-invariant PMFs, the passive tensor magnetic mode is similar in shape to the inflationary tensor mode. To address this, Zucca et al \cite{Zucca:2016iur} separately considered the cases with a fixed $r=0$, and when the two parameters are co-varied. It was found that this degeneracy does not affect the current bounds since the B-mode data on the relevant scales is still very weak. This will change in the future, when data from experiments such as the Simons Observatory \cite{Suzuki:2015zzg}, designed to detect $r \sim 0.001$, becomes available. Having a reliable B-mode spectrum measured over the $50 < l < 3000$ range would simultaneously constrain the amplitude and the generation epoch of the PMF, helping to narrow the search for the correct theory of magnetogenesis.

As Fig.~\ref{fig:PlanckContours} demonstrates, adding the B-mode data significantly tightens the constraints. In the case when the magnetic spectral index $n_B$ is allowed to vary, the 95\% CL bound on $B_{1 \, \rm{Mpc}}$ is reduced from $3.3$ nG to $1.55$ nG. The bounds are stronger when the spectral index is fixed to $n_B \approx -3$, which is of particular interest because the simplest models of inflationary magnetogenesis \cite{Turner:1987bw,Ratra:1991bn} predict a nearly scale-invariant PMF. In the case of $n_B=-2.9$, the 95\% CL bound is reduced from $2$ nG for Planck to $1.2$ nG for Planck+SPT. We note that the dependence on the smoothing scale disappears in the scale-invariant case, and $B_{1 \, \rm{Mpc}} = B_{\rm eff}$.

The other special cases is that of PMFs generated after inflation, {\it e.g.} during preheating \cite{DiazGil:2007dy,DiazGil:2008tf} or in phase transitions \cite{Vachaspati:1991nm}. Such fields have very small coherence lengths and are uncorrelated on cosmological scales. This leads to a generic prediction, based on causality, for such fields to have $n_B = 2$ on scales relevant to CMB anisotropies \cite{Jedamzik:1996wp,Durrer:2003ja,Jedamzik:2010cy}. Since these fields have most of their power concentrated on small scales, $1$ Mpc is not a representative scale for assessing their strength. It is more appropriate to use $\Omega_{B\gamma}$ or $B_{\rm eff}$, which quantify the total energy density in the PMF \cite{Kahniashvili:2009qi,Kahniashvili:2010wm,Pogosian:2011qv}. The Planck+SPT provides a constraint of $\Omega_{B\gamma} < 10^{-3}$ or $B_{\rm eff} < 100$ nG at 95\% CL, which is two orders of magnitude stronger than the BBN bound of $\Omega_{B\gamma} < 0.1$ \cite{Kernan:1995bz,Grasso:1996kk,Cheng:1996yi,Yamazaki:2012jd}. The corresponding bound on $B_{1 \, \rm{Mpc}}$ happens to be $B_{1 \, \rm{Mpc}}< 0.002$ nG at 95\% CL and represents the fact that most of the PMF power is concentrated on much smaller scales.

Future BB measurements will not seriously change the current  constraint on the PMF because of the quartic scaling of CMB spectra with the field strength. Instead, the mode-coupling induced by FR scales linearly and should offer a more sensitive probe. The method has already been applied to the POLARBEAR maps \cite{Ade:2015cao} to derive a $95$\% CL bound of $93$ nG on the scale-invariant PMF\footnote{Note that the POLARBEAR forecast performed in \cite{De:2013dra} assumed a frequency of $90$ GHz and $f_{\rm sky}=0.024$, while \cite{Ade:2015cao} used the actual $148$ GHz channel data with $f_{\rm sky}=0.0006$. This accounts for the significant difference between the $2\sigma$ bound of $3$ nG forecasted in \cite{De:2013dra} and the $93$ nG bound found in \cite{Ade:2015cao}.}. As we will see in the next subsection, the upcoming experiments will improve this bound by up to three orders of magnitude. Such a dramatic improvement will be due to a combination of factors, such as utilizing 100 GHz channels in addition to just the 150 GHz map used in \cite{Ade:2015cao}, having a larger sky fraction, $f_{\rm sky} \sim 0.2$ vs $0.0006$ used in \cite{Ade:2015cao}, and having lower noise levels, $\sigma_P \sim 2$ $\mu$K-arcmin vs the $6.9$ $\mu$K-arcmin in \cite{Ade:2015cao}. Thus, even though today the mode-coupling method is not competitive compared to using the B-mode spectrum, it holds a huge promise for the upcoming and future experiments.

\subsection{Forecasted bounds from SPT-3G, the Simons Observatory, CMB-S4 and Space Probe}

\begin{table*}[tbp]
\centering
\begin{tabular}{|c||c|c||c|c||c|c||c|c||c|c|}
\multicolumn{1}{c||}{} & \multicolumn{2}{c||}{PB \cite{Ade:2015cao}} &  \multicolumn{2}{|c||}{SPT-3G} & \multicolumn{2}{|c||}{Simons Obs.} & \multicolumn{2}{|c||}{CMB-S4} & \multicolumn{2}{|c}{Space Probe} \\
\multicolumn{1}{c||}{} &  \multicolumn{2}{c||}{$f_{sky}=0.0006$} &  \multicolumn{2}{|c||}{$f_{sky}=0.06$} & \multicolumn{2}{|c||}{$f_{sky}=0.3$} & \multicolumn{2}{|c||}{$f_{sky}=0.4$} & \multicolumn{2}{|c}{$f_{sky}=0.7$} \\
\hline
Freq & $\theta_{\rm fwhm}$ &  $\sigma_P$ & $\theta_{\rm fwhm}$ &  $\sigma_P$ & $\theta_{\rm fwhm}$ &  $\sigma_P$  & $\theta_{\rm fwhm}$ &  $\sigma_P$ & $\theta_{\rm fwhm}$  &  $\sigma_P$ \\
(GHz) &  ($'$)  & ($\mu$K-$'$) &  ($'$)  & ($\mu$K-$'$) &  ($'$)  &  ($\mu$K-$'$)  &  ($'$)  &  ($\mu$K-$'$) &  ($'$)  &  ($\mu$K-$'$) \\
\hline
             45 & - & - & - & - & - & - & - & - & $18$ & $8$ \\
             70 & - & - & - & - & - & - & - & - & $11$ & $4$ \\
100 & - & - & $2$ & $6$ & $5.2$ & $7$ & $1.5$ & $1.4$ & $8$ & $3$ \\
150 & 3.5 & 6.9 & $1.2$ & $3.5$ & $3.5$ & $6$ & $1$ & $1.4$ & $5$ & $2.5$ \\
220 & - & - & $1$ & $6$ & $2.7$ & $6$ & - & - & $3.5$ & $5$ \\
\hline
\end{tabular}
\caption{\label{tab:exp} Parameters of the future experiments considered in our forecast. Parameters of the $148$ GHz POLARBEAR (PB) channel used in a recent analysis of Faraday rotation in \cite{Ade:2015cao} are provided for comparison.}
\end{table*}

The SPT B-mode spectrum \cite{Keisler:2015hfa}, shown in Fig.~\ref{fig:VecBmode}, constrains the magnetic vector mode, which contributes roughly on the same angular scale as the weak lensing B-mode. Over the next few years, Stage III CMB experiments, such as SPT-3G \cite{Benson:2014qhw} and the Simons Observatory (SO) \cite{Suzuki:2015zzg}, will measure the B-mode spectrum with a significantly better accuracy. However, while they will reduce the uncertainty in the PMF contribution to BB by more than an order of magnitude \cite{Sutton:2017jgr}, it will translate into only a minor improvement in the bound on the PMF strength thanks to the quartic dependence of the spectrum on $B_\lambda$. It will remain at or just below $1$ nG even after the data from a future Stage IV ground based experiment \cite{Abazajian:2016yjj} or a future space mission become available. 

The FR induced mode-coupling, on the other hand, scales linearly with $B_\lambda$, and the qualitative improvements in the quality of CMB polarization maps will translate into dramatically improved bounds on nearly scale-invariant PMFs. As we show below, the $1\sigma$ uncertainty in $B_\lambda$ will decrease from about $\sim 50$ nG today \cite{Ade:2015cao} to $\sim 0.5$ nG for Stage III and to $\sim 0.1$ nG for Stage IV experiments. Crossing the $1$ nG threshold is significant for ruling out a purely primordial origin (requiring no dynamo action) of the galactic magnetic field, and the FR induced mode-coupling will achieve this milestone.

In what follows, we will present forecasts for three representative classes of experiments:
\begin{itemize}
\item Stage III ground based experiments, such as the planned SPT-3G \cite{Benson:2014qhw} and the Simons Observatory (SO) \cite{Suzuki:2015zzg}. Both are expected to measure the Stokes parameters at three frequencies: 95, 150 and 220 GHz. SPT-3G will offer $\sim 1$ arcmin resolution covering $6$\% of the sky and will achieve noise levels below the weak lensing B-mode signal ($< 5 \ \mu$K-arcmin). The SO will cover a larger fraction of the sky with a resolution of a few arcmin and noise levels comparable to the weak lensing B-mode.
\item A proposed Stage IV ground based experiment (CMB-S4) \cite{Abazajian:2016yjj}, that will cover 40\% of the sky at 100 and 150 GHz with $\sim 1$ arcmin resolution and noise levels of $\sim 1$ $\mu$K-arcmin.
\item A space mission, similar to the proposed CMBPol EPIC-IM \cite{Baumann:2008aq} and CORE \cite{Delabrouille:2017rct}, mapping the full sky using multiple channels covering a wide range of frequencies at a resolution of a few arcmin and noise levels comparable to the weak lensing B-mode. We will consider the channels in the 45-150 GHz range, where the mode-coupling induced by FR can be significant.
\end{itemize}
The parameters assumed for each of the experiments are summarized in Table~\ref{tab:exp}. Since the mode-coupling estimator is generally more sensitive to PMFs that have a large fraction of their power on large scales, we only consider the case of a scale-invariant PMF, with $n_B =-2.9$, with our conclusions being valid for a range of $n_B$ close to that value.

The forecasted bounds on the scale-invariant PMF amplitude are shown in Table~\ref{tab:forecasts} for different choices of parameters $f_L$ and $f_G$ quantifying the fraction of the weak lensing contribution to BB and the fraction of the galactic rotation measure, respectively. The case of no de-lensing corresponds to $f_L=1$, while a perfect de-lensing corresponds to $f_L=0$. Similarly, $f_G=1$ and $0$ correspond to no galaxy subtraction and a full subtraction, respectively. While a full subtraction of either weak lensing or the galactic rotation measure is impossible in practice, we opt to include these possibilities to demonstrate the ultimate sensitivity of the quadratic estimator to the PMF.

We find that the two Stage III experiments will have comparable sensitivities to the PMF, promising to achieve $\sim 0.6-0.7$ nG for SPT-3G and $\sim 0.4$ nG for the SO at $68$\% CL. While SPT-3G will offer a higher resolution and lower noise levels that the SO, it will cover a smaller fraction of the sky. Because the SPT-3G detector noise is expected to be securely below the weak lensing BB, the SPT-3G bounds are more sensitive to the degree of de-lensing. On the other hand, the expected noise levels of the SO are comparable to the weak lensing contribution and subtracting it does not make as much of a difference. Instead, the stronger bounds on the PMF from the SO are due to a larger sample of $ll'$ pairs available with the larger sky coverage. As expected, the superior accuracy with which Stage III experiments will measure the B-mode spectrum compared to the existing data from SPT will not lead to a notable improvement in the BB-based bound on the PMF strength. Instead, the mode-coupling based bound will be reduced by 2 orders of magnitude, from $\sim 50$ to $\sim 0.5$ nG.

The SO is considered to be a path-finder experiment towards a future Stage IV experiment. Both will cover about half the sky, but CMB-S4 will do it at a much higher resolution and lower noise levels. It can constrain the PMF to below $0.2$ nG without any de-lensing and can, in principle, achieve bounds below $0.1$ nG. The bounds would get even tighter if channels at frequencies lower than $95$ GHz were added. However, at some point, the galactic RM, which is roughly equivalent to the RM produced by a $\sim 0.1$ nG scale-invariant PMF \cite{De:2013dra}, will pose an obstacle to pushing the bound lower.

\begin{table*}[tbp]
\centering
\begin{tabular}{c|c||c|c||c|c||c|c||c|c||c|c|}
 \multicolumn{2}{c||}{} 
 & \multicolumn{2}{c||}{Current} &  \multicolumn{2}{|c||}{SPT-3G} & \multicolumn{2}{|c||}{SO} & \multicolumn{2}{|c||}{CMB-S4} & \multicolumn{2}{|c}{Space Probe} \\
 \multicolumn{2}{c||}{}  & \multicolumn{2}{c||}{(nG)} &  \multicolumn{2}{|c||}{(nG)} & \multicolumn{2}{|c||}{(nG)} & \multicolumn{2}{|c||}{(nG)} & \multicolumn{2}{|c}{(nG)} \\
\hline
$f_{L}$ & $f_{G}$ & FR$^{a}$
&  BB+$^{b}$
& FR &  BB$^{c}$
 & FR &  BB$^{c}$  & FR &  BB$^{c}$ & FR & BB$^{c}$ \\
\hline
\hline
1 & 1 & $50$ & $1$ & $0.75$ & $0.9$ & $0.45$ & $1$ 	& $0.18$  & $0.5$  	& $0.17$  & $0.7$  \\
1 & 0 & - & - & $0.75$  & - 		& $0.45$  & - 		& $0.14$  & - 		& $0.12$  & - \\
0 & 1 & - & - & $0.6$    & $0.8$  	& $0.37$  & $0.9$  	& $0.09$  & $0.4$  	& $0.14$  & $0.6$  \\
0 & 0 & - & - & $0.55$  & -  		& $0.37$  & - 		& $0.07$  & - 		& $0.09$  & - \\
\hline
\end{tabular}
\caption{\label{tab:forecasts} Current and forecasted $68$\% CL bounds on the amplitude $B_\lambda $ of a scale-invariant PMF for different choices of parameters $f_L$ and $f_G$ quantifying the fraction of the weak lensing contribution to BB and the fraction of the galactic rotation measure, respectively. {\footnotesize $^{a}$Estimated $68$\% CL bound based on the $95$\% CL bound of $93$ nG derived by POLARBEAR in \cite{Ade:2015cao}. $^{b}$The $68$\% CL bound derived in \cite{Zucca:2016iur} using TT, EE, TE spectra from Planck and BB from SPT. $^{c}$Based on the B-mode spectrum alone, assuming fixed cosmological parameters.}
} 
\end{table*}

As one can see from Table~\ref{tab:forecasts}, a realistic space mission would not necessarily offer stronger bounds than CMB-S4. A space based experiment would generally be better for studying large scale CMB anisotropies and for measurements at higher frequencies needed to better understand the galactic foregrounds. Since the mode-coupling estimator relies on measurements at $l\sim 1000$, it does not benefit significantly from a larger sky fraction. Also, the galactic rotation measure is very strong near the galactic plane, restricting the useful fraction of the sky to $f_{\rm sky} \sim 0.2$ around the galactic poles. Nevertheless, we find that a space probe such as EPIC-IM or CORE would offer bounds of $\sim 0.1$ nG, comparable to CMB-S4.

Overall, our forecasts show that one can achieve bounds close to $0.1$ nG without worrying about FR due to our galaxy. Going below $0.1$ nG would only be possible \cite{De:2013dra,Pogosian:2013dya} if there was an independent measurement of the galactic RM. A full sky RM map was constructed in Oppermann et al \cite{Oppermann:2011td} based on a compilation of RMs of known radio sources. Their map was used in \cite{De:2013dra}, as well as in the present analysis, to estimate the obstruction to the PMF posed by the galaxy. While the map of \cite{Oppermann:2011td} has large uncertainties, it is plausible to expect better maps, based on many more radio sources, to become available in the future. Thus, in principle, it should be possible to use the FR mode-coupling estimator to push the bound on the PMF below $0.1$ nG.

\section{Outlook}

Cosmology has entered the era of precision B-mode science thanks to the breakthrough results achieved by POLARBEAR \cite{Ade:2014afa}, BICEP/Keck \cite{Ade:2014xna,Array:2015xqh} and SPT \cite{Keisler:2015hfa}. The next generation of CMB polarization experiments will significantly improve the accuracy of B-mode measurements. In addition to searching for signatures of inflationary gravitational waves \cite{Crittenden:1993ni} and PMFs, they will probe neutrino masses \cite{Abazajian:2013oma}, modifications of gravity \cite{Amendola:2014wma,Raveri:2014eea}, cosmic (super)strings \cite{Seljak:1997ii,Avgoustidis:2011ax,Moss:2014cra,Lizarraga:2016onn} and other fundamental physics \cite{Abazajian:2016yjj}.
 
In this article, we have reviewed the most relevant B-mode signatures of a PMF. At present, and in the near term, the tightest bounds on the PMF strength will come from the vector and tensor mode magnetic contributions to the BB spectrum. When accurate measurements of BB at lower $l$ become available, it will be possible to simultaneously constrain the amplitude and the generation epoch of the PMF, which is of great relevance to theories of the early universe. Also, having accurate estimates of parity-odd TB and EB spectra would allow to constrain the helical component of the PMF\cite{Kahniashvili:2005xe,Kunze:2011bp,Ballardini:2014jta}, which has been neglected in the present work. 

In the future, looking for mode-coupling correlations induced by FR will produce tighter bounds. For the mode-coupling estimator approach to become competitive, one needs an experiment with a sufficiently high resolution to collect as many modes as possible near the peak of the EE spectrum, or in the $500<l<3000$ range. One also needs noise levels below the weak lensing B-mode, which is around $5$ $\mu$K-arcmin. Achieving a higher resolution and lower noise is, in principle, a matter of cost -- it can be accomplished by building larger telescopes with more detectors. While still challenging, it is arguably a simpler task than subtracting the large galactic foregrounds that obscure the primordial B-modes at lower $l$. FR estimators would also be aided by having more channels at frequencies below $100$ GHz, but they do not benefit from increasing the sky coverage beyond $20$\%, as the galactic FR becomes too strong away from the galactic poles. 

B-modes offer a relatively clean probe of the PMF, as there are relatively few other sources contributing on the relevant scales. The most relevant competing contribution to the BB spectrum comes from weak lensing, which can be partially subtracted using de-lensing techniques \cite{Hu:2001kj,Seljak:2003pn}. Fortunately, the mode-coupling correlations induced by lensing are of opposite parity to those induced by FR, and do not interfere with reconstruction of the FR angle. The lensing does, however, contribute to the variance of the quadratic estimator of FR, which will be important for Stage III and Stage IV experiments with noise levels below $5$ $\mu$K-arcmin. The frequency dependence of the FR signal also makes it distinct from other potential sources of birefringence, {\it e.g.} caused by pseudo-scalar fields \cite{Pospelov:2008gg}.

Our analysis did not account for instrumental systematic effects, such as the beam asymmetry and imperfect scanning of the sky, which can play an important role. Such systematics generate non-zero parity-odd spectra of EB and TB type, in addition to contributing to all parity-even spectra, including BB \cite{Shimon:2007au}. The impact of instrumental systematics on the ability to constrain the PMF is currently under investigation \cite{CPR_paper}.

The other important obstruction comes from the galactic RM, which can hinder the detection of FR from the PMF. To detect PMF strengths below $0.1$ nG one would need a more accurate galactic RM map \cite{Oppermann:2011td}. However, even with no de-lensing and no subtraction of the galactic RM, Stage III experiments will be able to decrease the upper bound on the PMF strength to $\sim 0.5$ nG, while a Stage IV experiment or an EPIC-IM class space mission would reduce it to below $0.2$ nG. This will conclusively rule out or confirm the existence of $\sim nG$ strength PMFs required to generate $\mu$G galactic fields with no help from a dynamo mechanism.

\acknowledgements

We benefited from earlier collaborations with Soma De, Yun Li, Bess Ng, Tanmay Vachaspati and Amit Yadav, as well as communications and discussion with Camille Bonvin, Fabio Finelli, Tina Kahniashvili, Brian Keating, Antony Lewis, Daniela Paoletti, Richard Shaw and Meir Shimon. We acknowledge support by the Natural Sciences and Engineering Research Council of Canada (NSERC). This research was enabled in part by support provided by WestGrid \cite{westgrid} and Compute Canada \cite{compute}.

\end{document}